\documentclass[conference,a4paper, onecolumn]{IEEEtran}

\usepackage{subfigure}
\usepackage{graphicx}
\usepackage{amssymb,amsmath,bm,bbm,booktabs}
\usepackage{units} % (1)
\usepackage{bbm,natbib}
\usepackage{enumerate}
\usepackage[normalem]{ulem}

\newtheorem{LEO}{LEO}
\newtheorem{PEO}{PEO}
\newtheorem{Lemma}[LEO]{Lemma}
\newtheorem{Proposition}[PEO]{Proposition}

\begin{document}

%\doi{10.1080/0266476YYxxxxxxxx}
%\jvol{03} \jnum{00} \jyear{2011} \jmonth{Month}

%\markboth{Frery et al.}{Chilean Journal of Statistics}

%\articletype{}

\def \tr {\mathop{\rm tr}\nolimits}
\def \etr {\mathop{\rm etr}\nolimits}
\def \diag {\mathop{\rm diag}\nolimits}
\def\build#1#2#3{\mathrel{\mathop{#1}\limits^{#2}_{#3}}}
\newcommand{\Half}{\mbox{$\frac{1}{2}$}}
\newcommand{\cuarto}{\mbox{$\frac{1}{4}$}}
\def \R{\mathop{\rm Re}\nolimits}
\newcommand {\findemo}{\hfill$\square$}

\title{Information Theory and Image Understanding: An Application to Polarimetric SAR Imagery}

\author{Frery, A.\ C.$^{\rm 1,\ast}$\thanks{$^\ast$Corresponding author. Email: acfrery@gmail.com
    \vspace{6pt}}, Nascimento, A.\ D.\ C.$^{\rm 2}$, and Cintra, R.\ J.$^{\rm 2}$\\\vspace{6pt}
    $^{\rm 1}${{Instituto de Computa\c{c}\~ao, Universidade Federal de Alagoas, Macei\'o, Brasil}}\\
    $^{\rm 2}${{Departamento de Estat\'istica, Universidade Federal de Pernambuco, Recife, Brazil}}
%    \\\vspace{6pt}
%    \received{Received: 00 Month 200x \sep  Accepted in final form: 00 Month 200x}
}

\maketitle

\begin{abstract}
%Quantifying the image contrast between different regions is central in image understanding.
This work presents a comprehensive examination of the use of information theory for understanding Polarimetric Synthetic Aperture Radar (PolSAR) images by means of contrast measures that can be used as test statistics.
Due to the phenomenon called `speckle', common to all images obtained with coherent illumination such as PolSAR imagery, accurate modelling is required in their processing and analysis.
The scaled multilook complex Wishart distribution has proven to be a successful approach for modelling radar backscatter from forest and pasture areas.
Classification, segmentation, and image analysis techniques which depend on this model have been devised, and many of them employ some kind of dissimilarity measure.
Specifically, we introduce statistical tests for analyzing contrast in such images. 
These tests are based on the chi-square, Kullback-Leibler, R\'enyi, Bhattacharyya, and Hellinger distances.
Results obtained by Monte Carlo experiments reveal the Kullback-Leibler distance as the best one with respect to the empirical test sizes under several situations which include pure and contaminated data. 
The proposed methodology was applied to actual data, obtained by an E-SAR sensor over surroundings of We\ss ling, Bavaria, Germany.\\
%
%\begin{classcode} %	94A17
%Primary 62Fxx	 \sep Secondary  94A17.
%\end{classcode}
%
\end{abstract}

\section{Introduction}

The aim of remote sensing is to capture and to analyze scenes of the Earth.
Among the remote sensing technologies, Polarimetric Synthetic Aperture Radar (PolSAR) has achieved a prominent position~\citep{LeePottier2009PolarimetricRadarImaging}.

In general, PolSAR data are the result of the following procedure: orthogonally polarized electromagnetic pulses are transmitted towards a target, and the returned echo is recorded with respect to each polarization~\citep{LopezMartinezFabregas2003}.
Such data are processed in order to generate images and, as a consequence of the coherent illumination, they are contaminated with fluctuations on its detected intensity called `speckle'.
Although speckle is a deterministic phenomenon since it is fully reproducible, it has the effect of a random noise.
These alterations can significantly degrade the perceived image quality, as much as the ability of extracting information from the data.

Defining a stochastic identity for modeling PolSAR image regions is an important pre-processing step~\citep{Conradsen2003}.
The scaled multilook complex Wishart distribution has been successfully employed as a statistical model for homogeneous regions in PolSAR imagery~\citep{Freryetal22011}.
Several statistical image processing techniques use this distribution for segmentation~\citep{BeaulieuTouzi2004}, classification~\citep{KerstenandLeeandAinsworth2005}, and boundary detection~\citep{Schou2003}, to name a few applications.

Any parametric approach requires parameter estimation.
The scaled multilook complex Wishart law is indexed by a scalar known as the number of looks and a Hermitian complex matrix.
These quantities can be estimated by a number of techniques, but a single scalar measure would be more useful when dealing with samples from images.
Such measure can be referred to as `contrast' if it provides means for discriminating different types of targets~\citep{Gambinietal:IJRS:06,GambiniandMejailandJacobo-BerllesandFrery,Goudail2004}.
Suitable measures of contrast not only provide useful information about the image scene but also assume a pivotal role in several image analysis procedures~\citep{Schou2003}.

Recent years have seen an increasing interest in adapting information-theoretic tools to image processing~\citep{Goudail2004}.
In particular, the concept of stochastic divergence~\citep{LieseVajda2006} has found applications in areas as diverse as image classification~\citep{PuigandGarcia2003}, cluster analysis~\citep{Mak1996}, and multinomial goodness-of-fit tests~\citep{zografosetal1990}.
Coherent polarimetric image processing has also benefited, since divergence measures can furnish methods for assessing segmentation algorithms~\citep{Schou2003}.
\cite{MorioRefregierGoudailFernandezDupuis2009} analyzed the Shannon entropy for the characterization of polarimetric imagery considering the circular multidimensional Gaussian distribution.
In a previous work~\citep{HypothesisTestingSpeckledDataStochasticDistances}, several parametric methods based on the class $(h,\phi)$-divergences were proposed and submitted to a comprehensive examination.

The aim of this work is to assess the contrast capability of hypothesis tests based on stochastic distances between statistical models for PolSAR images.
Analytic expressions for the $\chi^2$, Kullback-Leibler, R\'enyi (of order $\beta$), Bhattacharyya, and Hellinger distances between scaled multilook complex Wishart distributions in their most general way are derived.
Subsequently, such measures are penalized by coefficients which depend on the size of two distinct samples, leading to the proposal of new homogeneity tests.

The performance of these five new hypothesis tests is analyzed by means of their observed test sizes using Monte Carlo in several possible scenarios including pure and contaminated data. 
This methodology is assessed with real data obtained by the E-SAR sensor.

The remainder of this paper is organized as follows.  
In Section~\ref{ChJS:sec1}, we introduce the image statistical modeling for the polarimetric covariance matrix. 
The hypothesis testing method proposed by \cite{salicruetal1994} is adapted for dealing with Hermitian positive definite matrix models in Section~\ref{ChJS:sec2}.
Section~\ref{ChJS:sec3} presents computational results obtained from synthetic and actual data analysis.
The main conclusions are then summarized in Section~\ref{ChJS:sec4}.

\section{A model for polarimetric data: the scaled multilook complex Wishart distribution}~\label{ChJS:sec1}

The PolSAR processing results in a complex scattering matrix, which is defined by intensity and relative phase data. 
In strict terms such matrix has possibly four distinct complex elements, namely $S_\text{VV}$, $S_\text{VH}$, $S_\text{HV}$, and $S_\text{HH}$.
However, under the conditions of the reciprocity theorem~\citep{UlabyElachi1990}, the scattering matrix can be simplified to a three-component vector, since $S_\text{HV}=S_\text{VH}$.
Thus, we have a scattering vector
$$
\boldsymbol{s}=[S_\text{VV}\quad S_\text{VH}\quad S_\text{HH}]^{t},
$$
where $(\cdot)^t$ indicates vector transposition.
\cite{FreemananDurden} presented an important description of this three-component representation.
As discussed by~\citet{Goodmanb}, the multivariate complex Gaussian distribution can adequately model the statistical behaviour of $\boldsymbol{s}$.
This is called `single-look PolSAR data representation' and hereafter we assume that the scattering vector has the dimension $p$, i.e., $\boldsymbol{s}=[S_1\quad S_2\quad \ldots \quad S_p]^t$.

Polarimetric data is usually subjected to multilook processing in order to improve the signal-to-noise ratio.
To that end, Hermitian positive definite matrices are obtained by computing the mean over $N$ independent looks of the same scene.
This results in the sample covariance matrix $\boldsymbol{Z}$ given by  \citep{EstimationEquivalentNumberLooksSAR}
$$
\boldsymbol{Z}=\frac{1}{L}\displaystyle \sum_{i=1}^L \boldsymbol{s}_i \boldsymbol{s}_i^{\text{H}},
$$
where $L$ is the number of looks, $\boldsymbol{s}_i$, $i=1,2,\ldots,L$, and $(\cdot)^{\text{H}}$ represents the complex conjugate transposition.
The sample covariance matrix follows a scaled multilook complex Wishart distribution with parameters $\boldsymbol{\Sigma}$ and $L$ as parameters, characterized by the following probability density function:
\begin{equation}
f_{\boldsymbol{Z}}(\boldsymbol{Z}';\boldsymbol{\Sigma},L) = \frac{L^{pL}|\boldsymbol{Z}'|^{L-p}}{|\boldsymbol{\Sigma}|^L \Gamma_p(L)} \exp\{
-L\operatorname{tr}(\boldsymbol{\Sigma}^{-1}\boldsymbol{Z}')
\},
\label{eq:denswishart}
\end{equation}
where $\Gamma_p(L)=\pi^{p(p-1)/2}\prod_{i=0}^{p-1}\Gamma(L-i)$, $\Gamma(\cdot)$ is the gamma function, $\operatorname{tr}(\cdot)$ represents the trace operator, $|\cdot|$ denotes the determinant operator, $\boldsymbol{Z}'$ denotes a single-outcome of $\boldsymbol{Z}$, the covariance matrix of $\boldsymbol{Z}$ is given by
\begin{equation*}
\boldsymbol{\Sigma}=\operatorname{E}\{\boldsymbol{s}\boldsymbol{s}^{*}\} = {\left[\begin{array}{cccc}
\operatorname{E}\{S_1^{}S_1^{*}\} & \operatorname{E}\{S_1^{}S_2^{*}\} & \cdots &\operatorname{E}\{S_1^{}S_p^{*}\} \\
\operatorname{E}\{S_2^{}S_1^{*}\} & \operatorname{E}\{S_2^{}S_2^{*}\} & \cdots &\operatorname{E}\{S_2^{}S_p^{*}\} \\
\vdots &  \vdots & \ddots &\vdots \\
\operatorname{E}\{S_p^{}S_1^{*}\} & \operatorname{E}\{S_p^{}S_2^{*}\} & \cdots & \operatorname{E}\{S_p^{}S_p^{*}\} \end{array} \right]},
\end{equation*}
where $\operatorname{E}\{\cdot\}$ and $(\cdot)^{*}$ denote expectation and complex conjugation, respectively.
This distribution is denoted by $\boldsymbol{Z}\sim \mathcal W(L,\boldsymbol{\Sigma})$ and it satisfies $\operatorname{E}(\boldsymbol{Z})=\boldsymbol{\Sigma}$~\citep{EstimationEquivalentNumberLooksSAR}.

\citet{leeetal1994b} derived many marginal distributions of the $\mathcal W(L,\boldsymbol{\Sigma})$ law and their transformations.

\subsection{Interpretability of the parameters $\boldsymbol{\Sigma}$ and $L$ in PolSAR images}

Parameters $\boldsymbol{\Sigma}$ and $L$ possess physical meaning. 

\begin{figure}[htb]
\centering
\includegraphics[width=.6\linewidth]{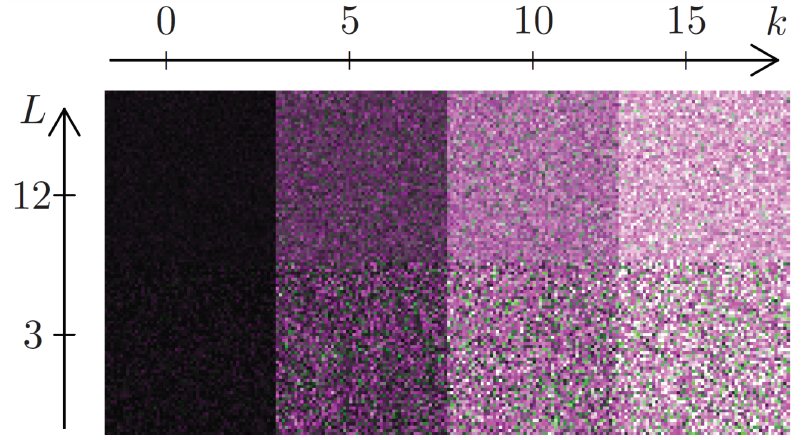}
\caption{Simulated PolSAR images following scaled multilook complex Wishart law with parameters $L=\{3,12\}$ and $k=\{0,5,10,15\}$ such that $\boldsymbol{\Sigma}=(1+k)B$.} 
\label{iluspolsar}
\end{figure}

The diagonal elements of $\boldsymbol{\Sigma}$ convey the brightness information of the respective channels. 
On its turn, an increasing number of looks $L$ implies better image signal-to-noise ratios.
Figure~\ref{iluspolsar} illustrates the influence of these parameters on simulated PolSAR images of size $100\times 100$ pixels.
Such synthetic images are generated from scaled multilook complex Wishart law with following parameters: $L\in\{3,12\}$ and $k\in\{0,5,10,15\}$ such that $\boldsymbol{\Sigma}=(1+k)B$, where 
\begin{equation}
B={\left[\begin{array}{ccc} 360932 & 11050+3759\textbf{i} & 63896+1581\textbf{i} \\ 
                & 98960 & 6593+6868\textbf{i} \\
                &  & 208843  \end{array} \right]},
\label{matrixEX}                
\end{equation}
where \textbf{i} is the imaginary unit.
Only the upper triangle and the diagonal are displayed because this covariance matrix $\boldsymbol{\Sigma}$ is Hermitian and, therefore, the remaining elements are the complex conjugates.
This matrix was observed in~\citep{PolarimetricSegmentationBSplinesMSSP} for representing PolSAR data of forested areas.
As shown in Figure~\ref{iluspolsar}, images with $L=12$ are less affected by speckle, and the brighter are the ones indexed by $k=15$, i.e., with lager determinants.
In this image, the HH (HV, VV, respectively) intensity was associated to the red (green, blue respectively) channel in order to compose a false color display.

\subsection{Maximum likelihood estimation under the Wishart model}

Let $\boldsymbol{Z}_k$ be a random matrix which follows a scaled multilook complex Wishart law with parameters $\boldsymbol{\gamma}=[L,\operatorname{vec}(\boldsymbol{\Sigma})^t]^t$, where $\operatorname{vec}(\cdot)$ is the column stacking vectorisation operator.
Its log-likelihood function is expressed by
$$
\ell_k(\boldsymbol{\gamma})=pL\log L+(L-p)\log |\boldsymbol{Z}_k| - L\log|\boldsymbol{\Sigma}|-\frac{p(p-1)}{2}\log\pi-\sum_{k=0}^{p-1}\log\Gamma(L-k)-L\operatorname{tr}(\boldsymbol{\Sigma}^{-1}\boldsymbol{Z}_k).
$$

According to \citet{HjorungnesandGesbert}, $\partial \ell_k(\boldsymbol{\gamma})/\partial \operatorname{vec}(\boldsymbol{\Sigma})=\operatorname{vec}(\partial \ell_k(\boldsymbol{\gamma})/\partial \boldsymbol{\Sigma})$ then, the score function based on $\boldsymbol{Z}_k$ is given by
$$
\nabla \ell_k(\boldsymbol{\gamma})=\left\{\begin{array}{l}
p(\log L+1)+\log|\boldsymbol{Z}_k|-\log|\boldsymbol{\Sigma}|-\sum_{i=0}^{p-1}\psi^{(0)}(L-i)-\operatorname{tr}\big(\boldsymbol{\Sigma}^{-1}\boldsymbol{Z}_k\big),\\
\\
L \operatorname{vec}\big(\boldsymbol{\Sigma}^{-1}\boldsymbol{Z}_k\boldsymbol{\Sigma}^{-1}-\boldsymbol{\Sigma}^{-1}\big).
\end{array}
\right.
$$

In the sequel we present the derivation of (a1)~the Hessian matrix $\mathcal J(\boldsymbol{\gamma})$, (a2)~the Fisher information matrix $\mathcal K(\boldsymbol{\gamma})$, and (a3)~the Cram\'er-Rao lower bound $\mathcal C(\boldsymbol{\gamma})$.
%To that end, the following term is pivotal: 
To that end, the following quantity plays a central role:
$$
\mathcal J_{\boldsymbol{\Sigma}\boldsymbol{\Sigma}}=\frac{\partial}{\partial \operatorname{vec}(\boldsymbol{\Sigma})^*}\Big\{\operatorname{vec}\Big(\frac{\partial \ell_k(\boldsymbol{\gamma})}{\partial \boldsymbol{\Sigma}}\Big)^t \Big\}.
$$

\cite{EstimationEquivalentNumberLooksSAR} showed that 
\begin{equation*}
T_1=-\frac{\partial \{\boldsymbol{\Sigma}^{-1}\boldsymbol{Z}\boldsymbol{\Sigma}^{-1}\}}{\partial \boldsymbol{\Sigma}}
=-\boldsymbol{\Sigma}^{-1} \otimes \boldsymbol{\Sigma}^{-1}\boldsymbol{Z}\boldsymbol{\Sigma}^{-1}-\boldsymbol{\Sigma}^{-1} \otimes \boldsymbol{\Sigma}^{-1}\boldsymbol{\Sigma}^{-1}\boldsymbol{Z},
\end{equation*}
where $\otimes$ denotes the Kronecker product.
Moreover, it is known that~\citep{HjorungnesandGesbert} $
T_2=\partial \boldsymbol{\Sigma}^{-1}/\partial \boldsymbol{\Sigma}= - \boldsymbol{\Sigma}^{-1} \otimes \boldsymbol{\Sigma}^{-1}$.
Thus, we have that
\begin{equation}\label{hessian}
\mathcal J_{\Sigma\Sigma}=L\{T_2-T_1\}=L\{\boldsymbol{\Sigma}^{-1} \otimes \boldsymbol{\Sigma}^{-1}
-\boldsymbol{\Sigma}^{-1} \otimes \boldsymbol{\Sigma}^{-1}\boldsymbol{Z}\boldsymbol{\Sigma}^{-1} 
-\boldsymbol{\Sigma}^{-1} \otimes \boldsymbol{\Sigma}^{-1}\boldsymbol{\Sigma}^{-1}\boldsymbol{Z}\}.
\end{equation}
From equation~\eqref{hessian}, it is possible to obtain that
$
\mathcal K_{\Sigma\Sigma}=\operatorname{E}\{-\mathcal J_{\Sigma\Sigma}\}=L \boldsymbol{\Sigma}^{-1} \otimes \boldsymbol{\Sigma}^{-1}$.
Thus, matrices $\mathcal J(\boldsymbol{\gamma})$,  $\mathcal K(\boldsymbol{\gamma})$, and $\mathcal C(\boldsymbol{\gamma})$ can be expressed as:
\begin{align}
\mathcal J(\boldsymbol{\gamma})&=\Bigl[ \begin{array}{cc}
\frac{p}{L} - \sum_{i=1}^{p-1} \psi^{(1)}(L-i)& \operatorname{vec}\big(\boldsymbol{\Sigma}^{-1}\boldsymbol{Z}\boldsymbol{\Sigma}^{-1}-\boldsymbol{\Sigma}^{-1}\big)^t \\
\operatorname{vec}\big(\boldsymbol{\Sigma}^{-1}\boldsymbol{Z}\boldsymbol{\Sigma}^{-1}-\boldsymbol{\Sigma}^{-1}\big)^* & \mathcal J_{\Sigma \Sigma} 
\end{array} \Bigr],\nonumber \\
\mathcal K(\boldsymbol{\gamma})&=\operatorname{E}\{-\mathcal J(\boldsymbol{\theta})\}= \Bigl[ \begin{array}{cc}
 \sum_{i=1}^{p-1} \psi^{(1)}(L-i) - \frac{p}{L}& \operatorname{vec}(\boldsymbol{0}_{p^2})^t \\
 \operatorname{vec}(\boldsymbol{0}_{p^2}) & L \boldsymbol{\Sigma}^{-1} \otimes \boldsymbol{\Sigma}^{-1} 
\end{array} \Bigr],
\label{derivedexpre2}\\
&\text{and}\nonumber \\
\mathcal C(\boldsymbol{\gamma})&=\mathcal K(\boldsymbol{\gamma})^{-1}=\Bigl[ \begin{array}{cc}
 (\sum_{i=1}^{p-1} \psi^{(1)}(L-i) - \frac{p}{L})^{-1}& \operatorname{vec}(\boldsymbol{0}_{p^2})^t \\
 \operatorname{vec}(\boldsymbol{0}_{p^2}) &  \boldsymbol{\Sigma} \otimes \boldsymbol{\Sigma}/L
\end{array} \Bigr],
\nonumber
\end{align}
respectively.

\cite{EstimationEquivalentNumberLooksSAR} derived the information Fisher matrix for the complex unscaled Wishart law, and found that the parameters of such distribution are not orthogonal.
However, dividing by the number of looks results in a block-diagonal Fisher information matrix as expressed by $\mathcal K(\boldsymbol{\gamma})$ in equation~\eqref{derivedexpre2}.
Thus, such scaling, whose density is given in equation~\eqref{eq:denswishart}, leads to a distribution with orthogonal parameters with good properties as, for instance, separable likelihood equations.
To the following, we discuss maximum likelihood (ML) estimation under such distribution.

Let $\{\boldsymbol{Z}_1,\boldsymbol{Z}_2,\ldots,\boldsymbol{Z}_N\}$ be a random sample of $\boldsymbol{Z}\sim \mathcal W(L,\boldsymbol{\Sigma})$ of size $N$, with $\boldsymbol{\theta}=(L,\boldsymbol{\Sigma})$ the parameter whose elements define the vector $\boldsymbol{\gamma}$, and let $\widehat{\boldsymbol{\theta}}=(\widehat{L},\widehat{\boldsymbol{\Sigma}})$ be its ML estimator. 
Expressing $N^{-1}\sum_{k=1}^N \nabla \ell_k(\widehat{\boldsymbol{\gamma}})=\boldsymbol{0}$, one has that
\begin{equation}
\left\{\begin{array}{l}
\widehat{\boldsymbol{\Sigma}}={N}^{-1}\sum_{k=1}^N {\boldsymbol{Z}_k},\\
\\
p\log \widehat{L} + {N}^{-1}\sum_{k=1}^N \log|\boldsymbol{Z}_k|-\log|\overline{\boldsymbol{Z}}|-\sum_{i=0}^{p-1}\psi^{(0)}(\widehat{L}-i)=0,
\label{eqscore1}
\end{array}\right.
\end{equation}
where $\widehat{L}$ is the ML estimator of the number of looks.
Thus, the ML estimators of $\boldsymbol{\Sigma}$ and $L$ are given by the sample mean and by the solution of equation~\eqref{eqscore1}, respectively.
The Newton-Raphson numerical optimization method was used for solving the latter, since a closed form solution is not trivially found.

\begin{figure}[htb]
\centering
\includegraphics[angle=0,width=.6\linewidth]{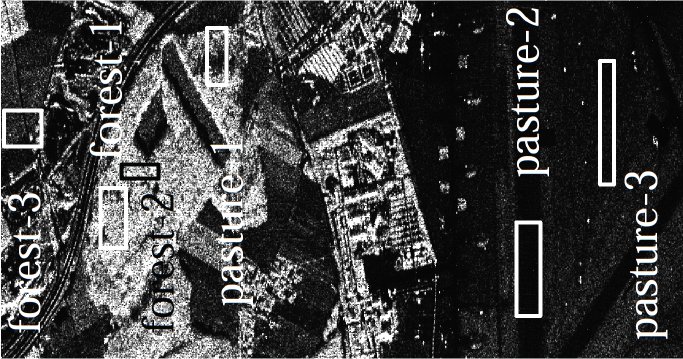}
\caption{E-SAR image (HH channel) with selected regions.} 
\label{fig0}
\end{figure}

In the following, this estimation is applied to actual data.
Figure~\ref{fig0} presents the HH channel of a polarimetric SAR image obtained by the E-SAR sensor over surroundings of We\ss ling, Germany.
The informed (nominal) number of looks is $3$.
The area exhibits two distinct types of target roughness: homogeneous (pasture) and heterogeneous (forest).
Table~\ref{tabelapplica} lists the ML estimates, as well as the sample sizes.

\begin{table}[hbt]
\centering   
\caption{Parameter estimates}\label{tabelapplica}
\begin{tabular}{c r@{.}l r@{.}l  cc} \toprule
Region & \multicolumn{2}{c}{$\widehat{L}$} & \multicolumn{2}{c}{$|\widehat{\boldsymbol{\Sigma}}|$}  & \# pixels  \\ 
\midrule
%\cmidrule(lr{.25em}){1-1} \cmidrule(lr{.25em}){2-3} \cmidrule(lr{.25em}){4-5} \cmidrule(lr{.25em}){6-6} 
pasture-1 &      2&870 &  7&934    &   2106   \\
pasture-2 &      2&573 & 74&660    &   2352   \\
pasture-3 &      2&889 & 26&452    &   3340    \\  
forest-1  &      2&638 & 33615&990 &   2870    \\ 
forest-2  &      2&727 & 10755&870 &   2496    \\ 
forest-3  &      2&303 &  7421&431 &   1020    \\ \bottomrule
\end{tabular}
\end{table}

\begin{figure}[htb]
\centering
\includegraphics[width=.8\linewidth]{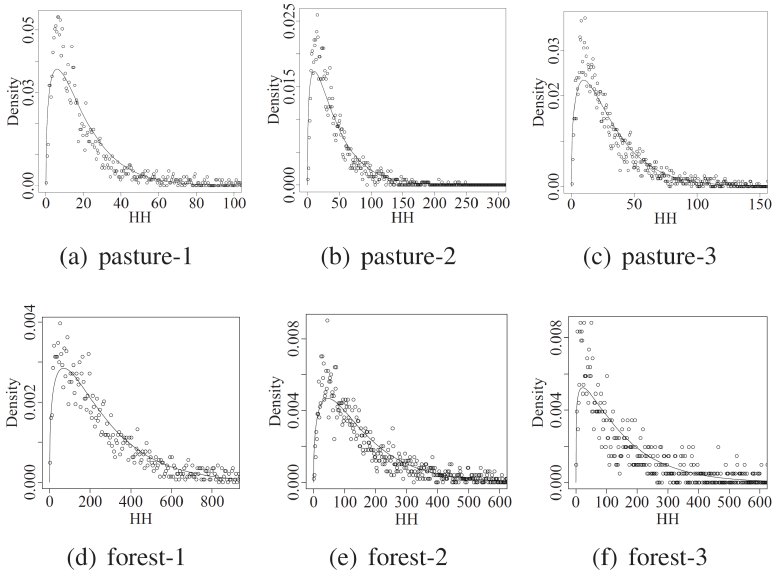}
\caption{Histograms and empirical densities of samples.} 
\label{fig1}
\end{figure}

Figure~\ref{fig1} depicts empirical densities of data samples from the selected forest and pasture regions.
Additionally, the associated fitted marginal densities are also shown for comparison.
In this case, the Wishart density collapses to gamma marginal densities, as demonstrated in~\citep{Hagedorn2006655}:
$$
f_{Z_i}(Z'_i;2\sigma^2_{i},L)=\frac{{Z'_i}^{L-1}}{2^L \sigma^{2L}_i\Gamma(L)}\exp\{-Z'_i/2\sigma^2_i\},
$$
where $\sigma^2_{i}$ is the element $(i,i)$ of $\boldsymbol{\Sigma}$, $p=3$, and $Z'_i$ is the entry $(i,i)$ of the random matrix $\boldsymbol{Z}$, and $1\leq i\leq 3$ with the association $1$ to HH, $2$ to HV and $3$ to VV.
%i\in\{1\text{ (HH)},2\text{ (HV)},3\text{ (VV)}\}$, }
%for $i\in\{HH,HV,VV\}$, where $\sigma^2_{i}$ is the element $(i,i)$ of $\boldsymbol{\Sigma}$, $p=3$, and $Z'_i$ is the entry $(i,i)$ of the random matrix $\boldsymbol{Z}$.  
The adequacy of the model to the data is noteworthy.
These samples will be used to validate our proposed methods in Section~\ref{ChJS:sec3}.

\section{Statistical information theory for random matrices}~\label{ChJS:sec2}

In the following we adhere to the convention that a (stochastic) `divergence' is any non-negative function of two probability measures.
If the function is also symmetric, it is called a (stochastic) `distance'.
Finally, we understand (stochastic) `metric' as a distance which satisfies the triangular inequality~\citep[chapters 1 and 14]{deza2009encyclopedia}.

An image can be understood as a set of regions formed by pixels which are observations of random variables following a certain distribution.
Therefore, stochastic dissimilarity measures could be used as features within image analysis techniques, since they may be able to assess the difference between the distributions that describe different image areas~\citep{HypothesisTestingSpeckledDataStochasticDistances}.
Dissimilarity measures were submitted to a systematic and comprehensive treatment in \citep{Ali1996,Csiszar1967,salicruetal1994} and, as a result, the class of $(h,\phi)$-divergences was proposed~\citep{salicruetal1994}.

Assume that $\boldsymbol{X}$ and $\boldsymbol{Y}$ are random matrices whose distributions are characterized by the densities $f_{\boldsymbol{X}}(\boldsymbol{Z}';\boldsymbol{\theta}_1)$ and $f_{\boldsymbol{Y}}(\boldsymbol{Z}';\boldsymbol{\theta}_2)$, respectively, where $\boldsymbol{\theta}_1$ and $\boldsymbol{\theta}_2$ are parameters.
Both densities are assumed to share a common support given by the cone of Hermitian positive definite matrices $\boldsymbol{\mathcal A}$.
The $(h,\phi)$-divergence between $f_{\boldsymbol{X}}$ and $f_{\boldsymbol{Y}}$ is defined by
\begin{eqnarray} \label{eq:eps2-no}
D_{\phi}^h(\boldsymbol{X},\boldsymbol{Y}) = 
h\biggl(\int_{\boldsymbol{\mathcal A}} \phi\Big( \frac{f_{\boldsymbol{X}}(\boldsymbol{Z}';\boldsymbol{\theta}_1)}{f_{\boldsymbol{Y}}(\boldsymbol{Z}';\boldsymbol{\theta}_2)}\Big) f_{\boldsymbol{Y}}(\boldsymbol{Z}';\boldsymbol{\theta}_2)\mathrm{d}\boldsymbol{Z}'\biggr),
\end{eqnarray}
where $\phi\colon (0,\infty)\rightarrow[0,\infty)$ is a convex function and $h\colon(0,\infty)\rightarrow[0,\infty)$ is a strictly increasing function with $h(0)=0$.
The differential element $\mathrm{d}\boldsymbol{Z}'$ is given by
$$
\mathrm{d}\boldsymbol{Z}'=\mathrm{d}Z'_{11}\mathrm{d}Z'_{22}\cdots\mathrm{d}Z'_{pp}\displaystyle\prod^p_{\underbrace{i,j=1}_{i<j}}\mathrm{d}\Re\{Z'_{ij}\} \mathrm{d}\Im\{Z'_{ij}\},
$$ 
where $Z'_{ij}$ is the $(i,j)$ entry of matrix $\boldsymbol{Z}'$, and operators $\Re\{\cdot\}$ and $\Im \{\cdot\}$ return real and imaginary parts of their arguments, respectively~\citep{Goodmanb}. 
If indeterminate forms appear in~\eqref{eq:eps2-no} due to the ratio of densities, they are assigned value zero.

Well-known divergences arise with adequate choices of $h$ and $\phi$. 
Among them, the following were examined: (i)~the $\chi^2$ divergence~\citep{Taneja2006}, (ii)~the Kullback-Leibler divergence~\citep{SeghouaneAmari2007}, (iii)~the R\'enyi divergence~\citep{Rachedetal2001}, (iv)~the Bhattacharyya distance~\citep{Kailath1967}, and (v)~the Hellinger distance~\citep{HypothesisTestingSpeckledDataStochasticDistances}.
As the triangular inequality is not necessarily satisfied, not every divergence measures is a metric~\citep{BurbeaRao1982}.
Additionally, the symmetry property is not followed by some of these divergence measures.
However, such tools are mathematically appropriated for comparing the distribution of random variables~\citep{Aviyente2003}.
Thus, the following expression has been suggested as a possible solution for this problem~\citep{SeghouaneAmari2007}:
$$
d_{\phi}^h(\boldsymbol{X},\boldsymbol{Y})=\frac{D_{\phi}^h(\boldsymbol{X},\boldsymbol{Y})+D_{\phi}^h(\boldsymbol{Y},\boldsymbol{X})}{2}.
$$

In analogy with~\citet{Goudail2004,InformationContrastAnalysisPolarimetric}, this paper defines distances as symmetrized versions of the divergence measures, i.e., 
a function $d_{\phi}^h:\boldsymbol{\mathcal A} \times \boldsymbol{\mathcal A} \rightarrow \mathbbm{R}$ is a distance on $\boldsymbol{\mathcal A}$ if, for all $\boldsymbol{X},\boldsymbol{Y}\in \boldsymbol{\mathcal A}$, the following properties holds:
\begin{enumerate}
\item Non-negativity: $d_{\phi}^h(\boldsymbol{X},\boldsymbol{Y})\geq 0$.
\item Symmetry: $d_{\phi}^h(\boldsymbol{X},\boldsymbol{Y})=d_{\phi}^h(\boldsymbol{Y},\boldsymbol{X})$.
\item Identity of indiscernibles: $d_{\phi}^h(\boldsymbol{X},\boldsymbol{Y})=0 \Leftrightarrow \boldsymbol{X}=\boldsymbol{Y} $.
\end{enumerate}
Thus, the  $\chi^2$, Kullback-Leibler, and R\'enyi divergences were made symmetric distances.
Table~\ref{tab-1} shows the functions $h$ and $\phi$ which lead to each of the above listed distances.

\begin{table}[hbt]
\centering   
\caption{($h,\phi$)-distance and related functions $\phi$ and~$h$}
\begin{tabular}{ccc}
\toprule
{ $(h,\phi)$-{distance}} & { $h(y)$} & { $\phi(x)$} \\
\midrule
$\chi^2$ &$y/4$& $(x-1)^2(x+1)/x$  \\ 
Kullback-Leibler & ${y}/{2}$ & $(x-1)\log x$  \\
R\'{e}nyi (order $\beta$) & $\frac{1}{\beta-1}\log((\beta-1)y+1),\;0\leq y<\frac{1}{1-\beta}$ & 
$\frac{x^{1-\beta}+x^{\beta}-\beta(x-1)-2}{2(\beta-1)},0<\beta<1$\\
Bhattacharyya  & $-\log(1-y),0\leq y<1$ & $-\sqrt{x}+\frac{x+1}{2}$ \\
Hellinger  & ${y}/{2},0\leq y<2$ &  $(\sqrt{x}-1)^2$  \\ \bottomrule
\end{tabular}
\label{tab-1}
\end{table}

In the following we discuss integral expressions of these $(h,\phi)$-distances. 
For simplicity, we suppress the explicit dependence on $\boldsymbol{Z}'$ and on the support $\boldsymbol{\mathcal A}$.
\begin{enumerate}[(i)]
\item The $\chi^2$ distance:
$$
d_{\chi^2}(\boldsymbol{X},\boldsymbol{Y})=\frac 12 [D_{\chi^2}(\boldsymbol{X},\boldsymbol{Y})+D_{\chi^2}(\boldsymbol{Y},\boldsymbol{X})]= \frac 12 \Big[   \int \frac{(f_{\boldsymbol{X}}-f_{\boldsymbol{Y}})^2}{2f_{\boldsymbol{X}}} + \int \frac{(f_{\boldsymbol{X}}-f_{\boldsymbol{Y}})^2}{2f_{\boldsymbol{Y}}}  \Big].
$$
The divergence $D_{\chi^2}$ has been used in many statistical contexts. 
For instance, \cite{BroniatowskiKeziou2009} proposed efficient hypothesis test combining it with the duality technique.

\item The Kullback-Leibler distance: 
\begin{align*} 
d_{\text{KL}}(\boldsymbol{X},\boldsymbol{Y})&=\frac 12 [D_{\text{KL}}(\boldsymbol{X},\boldsymbol{Y})+D_{\text{KL}}(\boldsymbol{Y},\boldsymbol{X})]
= \frac 12 \biggl[ \int f_{\boldsymbol{X}}\log{\frac{f_{\boldsymbol{X}}}{f_{\boldsymbol{Y}}}}+ \int f_{\boldsymbol{Y}}\log{\frac{f_{\boldsymbol{Y}}}{f_{\boldsymbol{X}}}} \biggl] \\
&=\frac{1}{2}\int(f_{\boldsymbol{X}}-f_{\boldsymbol{Y}})\log{\frac{f_{\boldsymbol{X}}}{f_{\boldsymbol{Y}}}}.
\end{align*}
The divergence $D_{\text{KL}}$ has a close relationship with the Neymman-Pearson lemma~\citep{est3} and its symmetrization has been suggested as a correction form of the Akaike information criterion, which is a descriptive measure for assessing the adequacy of statistical models.

\item The R\'{e}nyi distance of order $\beta$:
$$
\widetilde{d_{\text{R}}^{\beta }}(\boldsymbol{X},\boldsymbol{Y})= \frac 12 [ D_\text{R}^{\beta}(\boldsymbol{X},\boldsymbol{Y})+D_\text{R}^{\beta}(\boldsymbol{Y},\boldsymbol{X})]
=\frac{ \log\int f_{\boldsymbol{X}}^{\beta}f_{\boldsymbol{Y}}^{1-\beta} + \log\int f_{\boldsymbol{X}}^{1-\beta}f_{\boldsymbol{Y}}^{\beta}}{2(\beta-1)},
$$
where $0<\beta<1$.
The divergence $D_\text{R}^{\beta}$ has been used for analysing geometric characteristics with respect to probability laws~\citep{Andai2009777}.
By the Fejer inequality~\citep{Neuman1990}, we have that 
$$
{d_{\text{R}}^{\beta}}(\boldsymbol{X},\boldsymbol{Y}) \triangleq \frac{1}{\beta-1}\log   \frac{ \int f_{\boldsymbol{X}}^{\beta}f_{\boldsymbol{Y}}^{1-\beta}+ \int f_{\boldsymbol{X}}^{1-\beta}f_{\boldsymbol{Y}}^{\beta}}{2}
 \leq \widetilde{d_{\text{R}}^{\beta }}(\boldsymbol{X},\boldsymbol{Y}).
$$
Being ${d_{\text{R}}^{\beta}}$ more tractable than $\widetilde{d_{\text{R}}^{\beta}}$  for algebraic manipulation with the scaled multilook complex Wishart density, this paper will consider the former.  

\item [(iv)] The Bhattacharyya distance:
\begin{eqnarray*}
d_{\text{B}}(\boldsymbol{X},\boldsymbol{Y})= -\log \int\sqrt{f_{\boldsymbol{X}}f_{\boldsymbol{Y}}}.
\end{eqnarray*}
Goudail~\textit{et al.} showed that this distance is an efficient tool for contrast definition in algorithms for image processing~\citep{GoudailRefregierDelyon2004}.

\item The Hellinger distance:
\begin{eqnarray*} \label{eq:eps6}
d_{\text{H}}(\boldsymbol{X},\boldsymbol{Y})=1-\int\sqrt{f_{\boldsymbol{X}}f_{\boldsymbol{Y}}}.
\end{eqnarray*}
Estimation methods based on the minimization of $d_{\text{H}}$ have been successfully employed in the context of stochastic differential equations~\citep{Gietandmichel2008}.  
\end{enumerate}

When considering the distance between particular cases of the same distribution, only the parameters are relevant.
In this case, the parameters $\boldsymbol{\theta}_1$ and $\boldsymbol{\theta}_2$ replace the random variables $\boldsymbol{X}$ and $\boldsymbol{Y}$.
This notation is in agreement with that of \citet{salicruetal1994}.

In the following five subsections, we present the expressions of the discussed distances for the scaled multilook complex Wishart distribution, characterized by the density given in equation~\eqref{eq:denswishart} are presented.   

\subsection{Stochastic distances between scaled multilook complex Wishart laws}

In the following subsections,  analytic expressions for the stochastic distances $d_{\chi^2}$, $d_{\text{KL}}$, $d_\text{R}^\beta$, $d_{\text{B}}$, and   $d_{\text{H}}$ between two complex scaled multilook complex Wishart distributions are presented.
In all instances, the parameters $\boldsymbol{\theta}_X=(L_X,\boldsymbol{\Sigma}_X)$ and $\boldsymbol{\theta}_Y=(L_Y,\boldsymbol{\Sigma}_Y)$ were considered.
In order to avoid confusion with the determinant, the absolute value of the scalar $x$ will be denoted $\operatorname{abs}(x)$.

\subsubsection{the $\chi^2$ distance}
$$
d_{\chi^2}(\boldsymbol{\theta}_X,\boldsymbol{\theta}_Y)=
\frac{I_{XY}^{\chi^2}+I_{YX}^{\chi^2}-2}4,
$$ 
where 
%\begin{align*}% \bigr| |(2n_j\boldsymbol{\Sigma}_j^{-1}-n_i\boldsymbol{\Sigma}_i^{-1})^{-1}| \bigr|^{|2n_j-n_i|}
%I_{ij}^{\chi^2}=&\Big\{\frac{|\boldsymbol{\Sigma}_i|^{L_i}}{|\boldsymbol{\Sigma}_j|^{2L_j}} \bigr||(2L_j\boldsymbol{\Sigma}_j^{-1}-L_i\boldsymbol{\Sigma}_i^{-1})^{-1}|\bigr|^{|2L_j-L_i|}\Big\} \Big\{\frac{L_i^{-pL_i}}{L_j^{-2pL_j}} \prod_{k=1}^{p-1}\frac{(L_i-k)^k}{(L_j-k)^{2k}}(|2L_j-L_i-k|)^k\Big\}\\ 
%&\Big\{\frac{\Gamma(L_i-p+1)\Gamma(|2L_j-L_i-p+1|)}{\Gamma(L_j-p+1)^2}\Big\}^p,
%\end{align*}
\begin{align*}
I_{ij}^{\chi^2}=&\frac{|\boldsymbol{\Sigma}_i|^{L_i}}{|\boldsymbol{\Sigma}_j|^{2L_j}}\operatorname{abs}(|(2L_j\boldsymbol{\Sigma}_j^{-1}-L_i\boldsymbol{\Sigma}_i^{-1})^{-1}|)^{\operatorname{abs}(2L_j-L_i)} 
&\frac{L_i^{-pL_i}  }{L_j^{-2pL_j}  } \prod_{k=1}^{p-1}\frac{(L_i-k)^k}{(L_j-k)^{2k}}\operatorname{abs}(2L_j-L_i-k)^k \\
&\bigg[\frac{\Gamma(L_i-p+1)\Gamma(\operatorname{abs}(2L_j-L_i-p+1))}{\Gamma(L_j-p+1)^2}\bigg]^p
\end{align*}
for $(i,j)\in\{(X,Y),(Y,X)\}$.

\subsubsection{The Kullback-Leibler distance}

\begin{align*} 
d_{\text{KL}}(\boldsymbol{\theta}_X,\boldsymbol{\theta}_Y)=&\frac{L_{X}-L_{Y}}{2}\biggl\{\log\frac{|\boldsymbol{\Sigma}_{X}|}{|\boldsymbol{\Sigma}_{Y}|}
+p\bigl[\psi^{(0)}(L_{X}-p+1)-\psi^{(0)}(L_{Y}-p+1)\bigr]\nonumber \\
&-p\log\frac{L_X}{L_Y}+(L_Y-L_X)\sum_{i=1}^{p-1}\frac{i}{(L_X-i)(L_Y-i)}\biggr\}-\frac{p(L_{X}+L_{Y})}{2} \nonumber \\
&+\frac{\operatorname{tr}(L_{Y}\boldsymbol{\Sigma}_{Y}^{-1}\boldsymbol{\Sigma}_{X}+L_{X}\boldsymbol{\Sigma}_{X}^{-1}\boldsymbol{\Sigma}_{Y})}{2}.
%\label{dKL}
\end{align*}

\subsubsection{The R\'enyi distance of order $\beta$}

$$
d_{\text{R}}^{\beta}(\boldsymbol{\theta}_X,\boldsymbol{\theta}_Y)=\frac{1}{\beta-1}\log \frac{ I(\boldsymbol{\theta}_X,\boldsymbol{\theta}_Y)}2 ,
$$
such that
\begin{align*}
&I(\boldsymbol{\theta}_X,\boldsymbol{\theta}_Y)=\\
&\Bigl[\Gamma(L_{X}-p+1)^p |\boldsymbol{\Sigma}_{X}|^{L_{X}}L_{X}^{-pL_{X}} \displaystyle \prod_{i=1}^{p-1}(L_{X}-i)^i \Bigr]^{-\beta}
\Bigl[\Gamma(L_{Y}-p+1)^p |\boldsymbol{\Sigma}_{Y}|^{L_{Y}}L_{Y}^{-pL_{Y}} \displaystyle \prod_{i=1}^{p-1}(L_{Y}-i)^i \Bigr]^{\beta-1} \\
&\Gamma(E_1-p+1)^p  |\boldsymbol{\Sigma}_{XY}|^{E_1} \displaystyle \prod_{i=1}^{p-1}(E_1-i)^i  +\Bigl[(\Gamma(L_{X}-p+1)^p |\boldsymbol{\Sigma}_{X}|^{L_{X}}L_{X}^{-pL_{X}} \displaystyle \prod_{i=1}^{p-1}(L_{X}-i)^i \Bigr]^{\beta-1}\\
&\Bigl[\Gamma(L_{Y}-p+1)^p |\boldsymbol{\Sigma}_{Y}|^{L_{Y}}L_{Y}^{-pL_{Y}} \displaystyle \prod_{i=1}^{p-1}(L_{Y}-i)^i \Bigr]^{-\beta} 
\Gamma(E_2-p+1)^p  |\boldsymbol{\Sigma}_{YX}|^{E_2} \displaystyle \prod_{i=1}^{p-1}(E_2-i)^i ,
\end{align*}
where $E_1=\beta L_{X}+(1-\beta)L_{Y},\;E_2=\beta L_{Y}+(1-\beta)L_{X}$, $0<\beta<1$, and $\boldsymbol{\Sigma}_{ij}=|(L_i\beta \boldsymbol{\Sigma}_i^{-1}+L_j(1-\beta)\boldsymbol{\Sigma}_j^{-1})^{-1}|$
for $(i,j)\in\{(X,Y),(Y,X)\}$.

\subsubsection{The Bhattacharyya distance}

\begin{align*}
d_{\text{B}}(\boldsymbol{\theta}_X,\boldsymbol{\theta}_Y)&=\sum_{k=0}^{p-1}\log\frac{\sqrt{\Gamma(L_X-k)\Gamma(L_Y-k)}}{\Gamma(\frac{L_X+L_Y}{2}-k)}
+\frac{L_{X}\log|\boldsymbol{\Sigma}_X|}{2}+\frac{L_Y\log|\boldsymbol{\Sigma}_Y|}{2}\nonumber \\
&-\frac{p}{2}\bigl(L_{X}\log{L_{X}}+L_{Y}\log{L_{Y}}\bigr)-\frac{L_{X}+L_{Y}}{2}\log\biggl|\biggl(\frac{L_{X}\boldsymbol{\Sigma}_{X}^{-1}+L_{Y}\boldsymbol{\Sigma}_{Y}^{-1}}{2}\biggr)^{-1}\biggr|.
%\label{dbat}
\end{align*}

\subsubsection{The Hellinger distance}

$$
d_{\text{H}}(\boldsymbol{\theta}_X,\boldsymbol{\theta}_Y)=1-\frac{\Bigl|\Bigl(\frac{L_{X}\boldsymbol{\Sigma}_{X}^{-1}+L_{Y}\boldsymbol{\Sigma}_{Y}^{-1}}{2}\Bigr)^{-1}\Bigr|^{\frac{L_{X}+L_{Y}}{2}}}{|\boldsymbol{\Sigma}_{X}|^{\frac{L_{X}}{2}}|\boldsymbol{\Sigma}_{Y}|^{\frac{L_{Y}}{2}}}\sqrt{L_{X}^{pL_X} L_Y^{pL_Y}} 
\prod_{k=0}^{p-1}\frac{\Gamma\bigl(\frac{L_{X}+L_{Y}}{2}-k\bigr)}{\sqrt{\Gamma(L_X-k)\Gamma(L_Y-k)}}
$$

%\hskip+2ex

\subsubsection{Sensitivity analysis} 

Now we examine the behavior of the distances with respect to the variation of parameters.
First, we assume a fixed number of looks, namely $L=8$, and adopt the following parameters: $\boldsymbol{\theta}_1=(8,\boldsymbol{\Sigma}(360932))$ and $\boldsymbol{\theta}_2=(8,\boldsymbol{\Sigma}(x))$, where
$$\boldsymbol{\Sigma}(x)={\left[\begin{array}{ccc} x & 11050+3759\textbf{i} & 63896+1581\textbf{i} \\ 
                & 98960 & 6593+6868\textbf{i} \\
                &  & 208843  \end{array} \right]}.$$
As the covariance matrix is Hermitian, only the upper triangle and the diagonal are displayed.
The fixed covariance matrix $\boldsymbol{\Sigma}(360932)$ was observed by~\citet{PolarimetricSegmentationBSplinesMSSP} in PolSAR data of forested areas.

\begin{figure}[htb]
\centering
\subfigure[Varying $\boldsymbol{\Sigma} $\label{CurvesWishart2}]{\includegraphics[width=.48\linewidth,height=.68\linewidth]{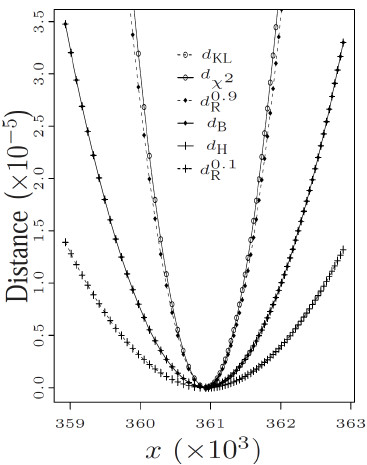}}
\subfigure[Varying $L$ \label{CurvesWishart1}]{\includegraphics[width=.48\linewidth,height=.68\linewidth]{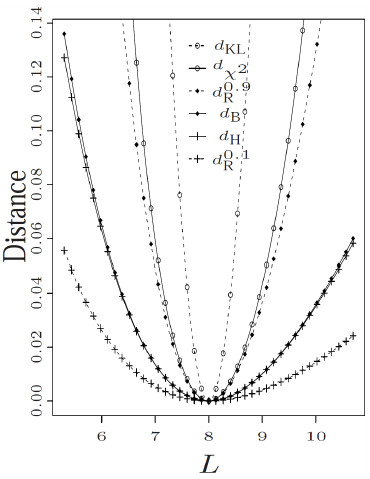}}
\caption{Sensitivity of the proposed distances.} 
\label{CurvesWishart}
\end{figure}

Figure~\ref{CurvesWishart2} shows the distances for $x \in [359000,363000]$.
The three uppermost curves, i.e., the ones that vary more abruptly, are $d_{\chi^2}$ and $d_\text{KL}$, which are overlapped, and $d_\text{R}^{0.9}$.
Analogously, distances $d_\text{B}$  and $d_\text{H}$ present roughly the same behavior and are also superimposed.
The least varying distance is $d_{\text{R}}^{0.1}$.

We also considered fixed covariance matrices with varying number of looks: $\boldsymbol{\theta}_1=(8,\boldsymbol{\Sigma}(360932))$ and $\boldsymbol{\theta}_2=(m,\boldsymbol{\Sigma}(360932))$, for $5\leq m \leq 11$.
Figure~\ref{CurvesWishart1} shows the distances, where we notice that $d_\text{KL}$ is the one that varies most, followed by $d_{\chi^2}$, $d_\text{R}^{0.9}$, $d_\text{B}$, $d_\text{H}$, and $d_\text{R}^{0.1}$.   
            
It is noteworthy that the distances are much more sensitive to variations of one element of the covariance matrix than to variations of the number of looks.

As presented by \cite{jieYuetal2008}, in order to make fair comparisons among stochastic distances, the distributions must be indexed by the same parameter. 
The evidence presented in Fig.~\ref{CurvesWishart} strongly suggests that $d_{\chi^2}$, $d_\text{KL}$, and $d_\text{R}^{0.9}$ outperform all other distances.
This is due to their higher sensitivity to variations around fixed parameter values, that makes them more suitable for discrimination purposes.

\subsection{Hypothesis test based on divergence for positive-definite Hermitian data}~\label{ChJS:sec2:ind}

In the following, the hypothesis test based on stochastic distances proposed by \cite{salicruetal1994} are recalled and applied to scaled multilook complex Wishart laws. 
  
Let $\widehat{\boldsymbol{\theta}}_1=(\widehat{\theta}_{11},\ldots,\widehat{\theta}_{1M})$ and $\widehat{\boldsymbol{\theta}}_2=(\widehat{\theta}_{21},\ldots,\widehat{\theta}_{2M})$ be the ML estimators of parameters $\boldsymbol{\theta}_1$ and $\boldsymbol{\theta}_2$ based on independent samples of size $N_X$ and $N_Y$, respectively.
Under the regularity conditions discussed by \citet[p. 380]{salicruetal1994} the following lemma holds:
\begin{Lemma}\label{prop-chi}
If $\frac{N_X}{N_X+N_Y} \xrightarrow[N_X,N_Y\rightarrow\infty]{} \lambda\in(0,1)$ and $\boldsymbol{\theta}_1=\boldsymbol{\theta}_2$, then
$$
S_{\phi}^h(\widehat{\boldsymbol{\theta}}_1,\widehat{\boldsymbol{\theta}}_2)=\frac{2 N_XN_Y}{N_X+N_Y}\frac{d^h_{\phi}(\widehat{\boldsymbol{\theta}}_1,\widehat{\boldsymbol{\theta}}_2)}{ h{'}(0) \phi{''}(1)}   \xrightarrow[N_X,N_Y\rightarrow\infty]{\mathcal D}\chi_{M}^2,
$$
where ``$\xrightarrow[]{\mathcal{D}}$'' denotes convergence in distribution. 
\end{Lemma}

Based on Lemma~\ref{prop-chi}, statistical hypothesis tests for the null hypothesis $\boldsymbol{\theta}_1=\boldsymbol{\theta}_2$ can be derived in the form of following proposition.
\begin{Proposition}\label{p-3}
Let $N_X$ and $N_Y$ be large and $S_{\phi}^h(\widehat{\boldsymbol{\theta}}_1,\widehat{\boldsymbol{\theta}}_2)=s$, then the null hypothesis $\boldsymbol{\theta}_1=\boldsymbol{\theta}_2$ can be re\-jec\-ted at le\-vel $\alpha$ if $\Pr( \chi^2_{M}>s)\leq \alpha$. 
\end{Proposition}

We denote the statistics based on the Kullback-Leibler, Bhattacharya, Hellinger, R\'enyi, and chi-square distances as $S_\text{KL}, S_\text{B}, S_\text{H}, S_\text{R}^{\beta}$, and $S_{\chi^2}$, respectively.

\section{Simulation and application}~\label{ChJS:sec3}

This section reports the assessment of the proposed methods as contrast measures.
Both synthetic and actual image data are considered.
Moreover, a robustness analysis was performed in order to characterize the behavior of such measures in the presence of outliers.

\subsection{Synthetic data analysis}

We assess the influence of estimation on the size of the new hypothesis tests using simulated data.
The study is conducted sampling from $L=L_X=L_Y\in\{4,8,16\}$ looks and from each of the covariance matrices observed in the forest areas presented in Figure~\ref{fig1}.
The sample sizes represent square windows of size $7\times 7$, $11\times 11$, and $20\times 20$ pixels, i.e., $N_X,N_Y \in\{49,121,400\}$, which are typical in image processing and analysis.
Nominal significance levels  $\alpha\in\{1\%,5\%\}$ are verified.

Let $T$ be the number of Monte Carlo replicas and $C$ the number of cases for which the null hypothesis is rejected at nominal level $\alpha$ when samples come from the same distribution.
The empirical test size is given by $\widehat{\alpha}_{1-\alpha}={C}/{T}$ .
Following the methodology described by~\citet{HypothesisTestingSpeckledDataStochasticDistances}, we used $T=5500$ replicas.

Table~\ref{table1} presents the empirical test size at $1\%$ and $5\%$ nominal levels, the execution time in miliseconds\footnote{All experiments were performed on a PC with an Intel Core 2 Duo processor \unit[2.10]{GHz}, \unit[4]{GB} of RAM, Windows XP and R~2.8.1}, and the distances mean ($\overline{d}$) and coefficient of variation ($\text{CV}$).  
The smallest empirical size and distance mean are in boldface.

%\begin{landscape}
\begin{table}[hbt]
\centering 
\scriptsize                                                                        
\caption{Empirical sizes for forest-1}\label{table1} 
\begin{tabular}{c@{ }r@{ }r  r@{\quad }r@{\,\,\,}c@{\,\,\,}r@{\,\,\,}r  r@{\quad }r@{\,\,\,}c@{\,\,\,}r@{\,\,\,}r r@{\quad}r@{\,\,\,}c@{\,\,\,}r@{\,\,\,}r}
\toprule 
\multicolumn{3}{c}{Factors}& \multicolumn{5}{c}{$L=4$} & \multicolumn{5}{c}{$L=8$} & \multicolumn{5}{c}{$L=16$} \\ 
\cmidrule(lr{.25em}){4-8} \cmidrule(lr{.25em}){9-13} \cmidrule(lr{.25em}){14-18} 
$S_{\phi}^h$ & $N_X$ & $N_Y$ & \rotatebox{90}{$1\%$} & \rotatebox{90}{$5\%$} & \rotatebox{90}{time (ms)} & \rotatebox{90}{$\overline{d}$} &  \rotatebox{90}{$\text{CV}$}
& \rotatebox{90}{$1\%$} & \rotatebox{90}{$5\%$} & \rotatebox{90}{time (ms)} & \rotatebox{90}{$\overline{d}$} &  \rotatebox{90}{$\text{CV}$} 
& \rotatebox{90}{$1\%$} & \rotatebox{90}{$5\%$} & \rotatebox{90}{time (ms)} & \rotatebox{90}{$\overline{d}$} &  \rotatebox{90}{$\text{CV}$} \\ 
\cmidrule(lr{.25em}){1-3} \cmidrule(lr{.25em}){4-8} \cmidrule(lr{.25em}){9-13} \cmidrule(lr{.25em}){14-18} 
$S_{\chi^2}$                                                                                                                                                              &49& 49 &  84.96	&	91.55	&	1.03	&	64.58	&	159.51	&	81.55	&	89.13	&	1.08	&	52.62	&	81.55	&	\textbf{78.69}	&	\textbf{87.05}	      &	1.09	&	{\textbf{48.70}}	&	78.16 \\
&49&121&  82.44	&	90.62	  &	1.11	&	50.74	&	86.63	  &	78.98	&	88.07	&	1.08	&	44.90	&	68.93	&	\textbf{76.64}	&	\textbf{86.64}	      &	1.07	&	{\textbf{42.26}}	&	65.09 \\
&49&400&  81.84	&	90.38	  &	1.02	&	47.40	&	66.63	  &	77.09	&	87.64	&	1.09	&	42.84	&	63.52	&	\textbf{74.93}	&	\textbf{85.11}	      &	1.06	&	{\textbf{40.71}}	&	61.99 \\
&121&121&  79.96	&	89.27	&	1.07	&	43.89	&	60.13	  &	76.16	&	86.58	&	1.09	&	40.16	&	59.26	&	\textbf{73.76}	&	\textbf{85.31}	      &	1.04	&	{\textbf{38.12}}	&	57.44 \\
&121&400&  78.78	&	88.51	&	1.12	&	41.76	&	56.54	  &	74.25	&	85.53	&	1.08	&	{\textbf{37.16}}&	54.32	&	\textbf{73.33}&	\textbf{84.84}&	1.12	&	{37.26} &	56.76 \\
&400&400&  77.96	&	87.60	&	1.09	&	38.88	&	52.01	  &	73.36	&	84.71	&	1.05	&	35.95	&	51.72	&	\textbf{71.51}	&	\textbf{83.56}	      &	1.10	&	{\textbf{34.99}} &	51.39 
\\
\cmidrule(lr{.25em}){1-3} \cmidrule(lr{.25em}){4-8} \cmidrule(lr{.25em}){9-13} \cmidrule(lr{.25em}){14-18}                                                               

$S_{\text{KL}}$                                                                                                                                                                                           
&49& 49 &   1.91	&	 7.18	&	0.44	&	10.59	&	45.65	  &	\textbf{1.05}	&	  \textbf{4.85}	&	0.49	&	 9.75	&	46.78	&	 0.91	        &	 4.22	        &	0.39	&	 {\textbf{9.39}}	&	47.80 \\                      
&49& 121 &   1.58	&	 6.35	&	0.41	&	10.47	&	45.78	  &	\textbf{1.00}	&	  \textbf{4.42}	&	0.45	&	 9.60 &	46.02	&	 0.78	        &	 3.67	        &	0.45	&	 {\textbf{9.24}} &	46.92  \\                     
&49& 400 &   1.56	&	 6.95	&	0.48	&	10.57	&	45.02	  &	\textbf{1.04}	&	  \textbf{4.35}	&	0.46	&	 9.67 &	46.36	&	 0.82	        &	 3.95	        &	0.47	&	 {\textbf{9.32}} &	47.49  \\                     
&121&121&   1.67	&	 6.62	&	0.43	&	10.56	&	45.79	  &	\textbf{0.85}	&	 \textbf{4.60}	&	0.49	&	9.66  &	46.80	&	 0.75	        &	3.78	        &	0.48	&	{\textbf{9.21}}  &	47.09  \\                     
&121&400&   1.82	&	 7.64	&	0.50	&	10.69	&	45.69	  &	0.76	        &	4.27	          &	0.46	&	9.49  &	45.90	&	\textbf{1.16}	&	\textbf{4.49}	&	0.47	&	{\textbf{9.43}}  &	48.43 \\                      
&400&400&   1.47	&	 6.91	&	0.42	&	10.55	&	44.91	  &	\textbf{1.00}	&	\textbf{4.16}	  &	0.50	&	9.58  &	46.21	&	0.58	        &	3.56	        &	0.46	&	{\textbf{9.26}}  &	46.46 \\                      
\cmidrule(lr{.25em}){1-3} \cmidrule(lr{.25em}){4-8} \cmidrule(lr{.25em}){9-13} \cmidrule(lr{.25em}){14-18}                                                                
                                 
$S_{\text{R}}^{\beta}$                                                                                                                                                    &49& 49 &   5.89	&	14.76	&	1.10	&	12.38	&	49.66	  &	\textbf{4.13}	&	12.93	        &	1.12	&	11.72	          &	50.22	 &	 4.20	         &	\textbf{11.40}	&	1.07	&	{\textbf{11.35}}	&	51.15 \\        
&49&121&    4.78	&	14.38	&	1.00	&	12.18	&	49.79	  &	3.89	        &	11.58	        &	1.10	&	11.52	          &	49.94	 &	\textbf{3.62}	 &	\textbf{10.36}	&	1.09	&	{\textbf{11.16}}	&	50.18 \\        
&49&400&    5.71	&	15.18	&	1.08	&	12.31	&	49.25	  &	4.25	        &	12.47	        &	0.96	&	11.62	          &	50.38	 &	\textbf{3.85}	 &	\textbf{11.11}	&	1.11	&	{\textbf{11.27}}	&	50.84 \\        
&121&121&   5.56	&	14.69	&	1.11	&	12.32	&	49.56	  &	4.07	        &	12.02	        &	0.99	&	11.59	          &	50.65	 &	\textbf{3.45}	 &	\textbf{10.71}	&	1.12	&	{\textbf{11.17}}	&	50.14 \\        
&121&400&   6.29	&	16.00	&	1.11	&	12.47	&	50.31	  &	\textbf{3.65}	&	\textbf{11.20}&	1.06	&	{\textbf{11.33}}&	49.50	 &	3.98	         &	11.36	          &	1.01	  &	11.40 &	51.57 \\                  
&400&400&   5.71	&	15.24	&	0.99	&	12.35	&	49.47	  &	3.69	        &	11.62	        &	1.13	&	11.50	          &	49.55	 &	\textbf{3.27}	 &	\textbf{11.35}  &	1.20	&	{\textbf{11.23}} &	49.49 \\        
\cmidrule(lr{.25em}){1-3} \cmidrule(lr{.25em}){4-8} \cmidrule(lr{.25em}){9-13} \cmidrule(lr{.25em}){14-18}                                                               
$S_{\text{B}}$                                                                                                                                                            
&49& 49 &   5.51	&	14.25	&	0.54	&	12.27	&	49.22	  &	 \textbf{3.96}	&	12.65	        &	0.51	&	11.66	           &	50.01	&	 4.07	        &	\textbf{11.31}	&	0.58	&	{\textbf{11.33}}	&	51.04 \\        
&49&121&   4.56	&	14.04	  &	0.65	&	12.10	&	49.48	  &	 3.78	          &	11.35	        &	0.51	&	11.48	           &	49.80	&	 \textbf{3.58}&	\textbf{10.29}	&	0.57	&	{\textbf{11.14}}	&	50.11 \\        
&49&400&   5.45	&	14.84	  &	0.56	&	12.25	&	49.01	  &	 4.18	          &	12.36	        &	0.57	&	11.59	           &	50.26	&	\textbf{3.84}	&	\textbf{11.02}	&	0.58	&	{\textbf{11.26}}	&	50.78 \\        
&121&121&   5.42	&	14.47	&	0.57	&	12.27	&	49.39	  &	 3.98	          &	11.95	        &	0.61	&	11.56	           &	50.56	&	\textbf{3.45}	&	\textbf{10.64}	&	0.56	&	{\textbf{11.16}}	&	50.09 \\        
&121&400&   6.20	&	15.75	&	0.56	&	12.44	&	50.20	  &	\textbf{3.62}	  &	\textbf{11.09}&	0.55	&	{\textbf{11.32}} &	49.44	&	3.98	        &	11.36	          &	0.51	&	11.40 &	51.54 \\                    
&400&400&   5.65	&	15.18	&	0.57	&	12.33	&	49.42	  &	3.67	          &	11.58	        &	0.57	&	11.50	           &	49.52	&	\textbf{3.27}	&	\textbf{11.33}	&	0.65	&	{\textbf{11.23}} &	49.47  \\       
\cmidrule(lr{.25em}){1-3} \cmidrule(lr{.25em}){4-8} \cmidrule(lr{.25em}){9-13} \cmidrule(lr{.25em}){14-18}                                                                $S_{\text{H}}$                                                                                                                                                            &49& 49 &   4.20	&	11.93	&	0.53	&	11.80	&	47.24	  &	 2.93	        &	10.36	        &	0.56	&	11.24	          &	48.11	&	 \textbf{2.78}	&	 \textbf{9.11}	&	0.54	&	{\textbf{10.93}}	&	49.10 \\          
&49&121 &   3.73	&	12.29	&	0.50	&	11.78	&	48.01	  &	 3.00	        &	10.05	        &	0.52	&	11.20	          &	48.41	&	 \textbf{2.87}	&	 \textbf{9.27}	&	0.60	&	{\textbf{10.87}}	& 48.80 \\          
&49&400 &   4.49	&	13.38	&	0.64	&	11.99	&	47.90	  &	 3.56	        &	10.93	        &	0.62	&	11.36	          &	49.16	&	 \textbf{3.27}	&	 \textbf{9.71}	&	0.55	&	{\textbf{11.04}}	&	49.72 \\          
&121&121&   4.89	&	13.58	&	0.58	&	12.08	&	48.56	  &	 3.60	        &	11.35	        &	0.62	&	11.39	          &	49.77	&	 \textbf{2.96}	&	 \textbf{9.87}	&	0.62	&	{\textbf{11.00}}	&	49.34 \\          
&121&400&   5.67	&	15.18	&	0.60	&	12.31	&	49.64	  &	 \textbf{3.35}&	\textbf{10.58}&	0.54	&	{\textbf{11.22}}&	48.96	&	 3.71	          &	10.80	          &	0.60	&	11.29 &	50.99 \\                      
&400&400&   5.40	&	14.93	&	0.55	&	12.27	&	49.17	  &	 3.51	        &	11.38	        &	0.51	&	11.44	          &	49.29	&	 \textbf{3.11}	&	\textbf{11.02}	&	0.58	&	{\textbf{11.18}} &	49.26 \\          
\bottomrule                      
\end{tabular}
\end{table}
%\end{landscape}

Average distances reduce when the number of looks increases, i.e., improving image quality corrects the statistics in terms of test size.
With the exception of the $\chi^2$ distance, these distances vary in the interval ($45.02,51.54$) regardless the sample sizes.
Fixing the sample sizes while varying the number of looks $L$, the test sizes obey the inequalities $S_{\text{KL}}\leq S_{\text{H}} \leq S_{\text{B}} \leq S_{\text{R}}^{\beta} \leq S_{\chi^2}$.

The test based on the Kullback-Leibler distance presented the best performance in terms of both empirical test size and execution time.
The other tests showed poorer performance, even with large samples and high number of looks.
Small samples or number of looks yield poor tests, and the test based on the $\chi^2$ distance is unacceptable.

\subsection{Image data analysis}

The methodology for assessing test size presented in Section~\ref{ChJS:sec2:ind} was applied to the three forest samples of the E-SAR image presented in Figure~\ref{fig0}.
Each sample was submitted to the following procedure~\citep{HypothesisTestingSpeckledDataStochasticDistances}: 
\begin{enumerate}[(b1)]
\item partition the sample in disjoint blocks of size $N_X$;
\item for each block from~(b1), split the remaining sample in disjoint blocks of size $N_Y$;
\item perform the hypothesis test as described in Proposition~\ref{p-3} for each pair of samples with sizes $N_X$ and $N_Y$.
\end{enumerate}

Table~\ref{table2} presents the results. 
Except for the test based on the $\chi^2$ distance, all test sizes were smaller than the nominal level; i.e., the tests did not reject the null hypothesis when it is true.
Although $S_{\chi^2}$ presented the worst performance in general, it is equal to zero when $N_X=N_Y=400$, showing the importance of the sample size on the test size.

%\begin{landscape}
\begin{table}[hbt]
\centering
\footnotesize
\caption{Empirical sizes for forests}\label{table2}
%\begin{tabular}{*{3}{p{6pt}}rrrr  rrrr  rrrr}
\begin{tabular}{c@{ }r@{ }r  r@{\,\,\,}r@{\,\,\,}r@{\,\,\,}r  r@{\,\,\,}r@{\,\,\,}r@{\,\,\,}r r@{\,\,\,}r@{\,\,\,}r@{\,\,\,}r}
\toprule
\multicolumn{3}{c}{Factors}& \multicolumn{4}{c}{forest-1} & \multicolumn{4}{c}{forest-2} & \multicolumn{4}{c}{forest-3} \\
 \cmidrule(lr{.25em}){1-3} \cmidrule(lr{.25em}){4-7} \cmidrule(lr{.25em}){8-11} \cmidrule(lr{.25em}){12-15}

$S_{\phi}^h$&$N_X$&$N_Y$ & \multicolumn{1}{c}{\rotatebox{90}{$1\%$}} & \multicolumn{1}{c}{\rotatebox{90}{$5\%$}} & \multicolumn{1}{c}{\rotatebox{90}{$\overline{d}\;(\times 10^{-1})$}} & \multicolumn{1}{c}{\rotatebox{90}{CV}}
& \multicolumn{1}{c}{\rotatebox{90}{$1\%$}} & \multicolumn{1}{c}{\rotatebox{90}{$5\%$}} & \multicolumn{1}{c}{\rotatebox{90}{$\overline{d}\;(\times 10^{-1})$}}& \multicolumn{1}{c}{\rotatebox{90}{CV}}
& \multicolumn{1}{c}{\rotatebox{90}{$1\%$}} & \multicolumn{1}{c}{\rotatebox{90}{$5\%$}} & \multicolumn{1}{c}{\rotatebox{90}{$\overline{d}\;(\times 10^{-1})$}}& \multicolumn{1}{c}{\rotatebox{90}{CV}}
\\ \cmidrule(lr{.25em}){1-3} \cmidrule(lr{.25em}){4-7} \cmidrule(lr{.25em}){8-11} \cmidrule(lr{.25em}){12-15}  

$S_{\chi^2}$
&49& 49 & 49.30 & 53.06 & $\text{10}^{\text{25}}$ &4065.70& 55.27 & 58.37 & $\text{10}^{\text{37}}$ & 4000.00 & 61.05 & 63.16 & $\approx\text{10}^{\text{47}}$&1378.40  \\     
&49& 121& 37.77 & 40.06 & $\text{10}^{\text{17}}$ &1565.32& 43.76 & 46.42 & $\text{10}^{\text{19}}$ & 2169.27 & 51.40 & 52.78 & $\text{10}^{\text{7}}\;$ & 642.51   \\    
&49& 400& 27.12 & 29.38 & $\text{10}^{\text{11}}$ &1330.26& 29.77 & 29.77 & $\text{10}^{\text{27}}$ & 1144.55 & 43.75 & 43.75 & $\text{10}^{\text{9}}\;$ & 396.99   \\    
&121&121& 14.62 & 16.60 & $\text{10}^{\text{7}}\;$ &1197.95& 24.21 & 30.00 & $\text{10}^{\text{21}}$ & 1378.40 & 46.43 & 50.00 & $\text{10}^{\text{8}}\;$ & 446.32   \\   
&121&400&  5.80 & 5.80  & 137.97                  &603.05 & 7.84  & 11.76 &68.36                    & 169.33  & 40.00 & 40.00 &494.91 & 179.09   \\                     
&400&400&  0.00 & 0.00  &  12.16                  &77.12  & 0.00  & 0.00  & 2.74                    & 39.25   & 0.00  &  0.00 &  6.54 & \multicolumn{1}{c}{$- $} \\     

\cmidrule(lr{.25em}){1-3} \cmidrule(lr{.25em}){4-7} \cmidrule(lr{.25em}){8-11} \cmidrule(lr{.25em}){12-15}                                                              
$S_{\text{KL}}$                                                                                                                                                         
&49& 49 & 0.00 & 0.00 &17.63 & 56.82& 0.00 & 0.00 & 15.63  &  47.24 &     0.00 &  0.00 & 46.73 &  57.31\\                                                               
&49& 121& 0.00 & 0.00 &10.91 & 56.50& 0.00 & 0.00 & 10.45  &  47.29 &     0.00 &  0.00 & 30.12 &  62.51    \\                                                           
&49& 400& 0.00 & 0.00 &7.81 & 60.15 & 0.00 & 0.00 &  7.47  &  46.14 &     0.00 &  0.00 & 18.07 &  73.62    \\                                                           
&121&121& 0.00 & 0.00 &7.09 & 44.14 & 0.00 & 0.00 &  8.02  &  54.38 &     0.00 &  0.00 & 29.83 &  53.86    \\                                                           
&121&400& 0.00 & 0.00 &4.27 & 47.16 & 0.00 & 0.00 &  5.28  &  60.25 &     0.00 &  0.00 & 8.15  &  52.58    \\                                                           
&400&400& 0.00 & 0.00 &2.95 & 38.16 & 0.00 & 0.00 & 16.99  &  59.64 &     0.00 &  0.00 & 2.61  &  \multicolumn{1}{c}{$- $} \\                                           
\cmidrule(lr{.25em}){1-3} \cmidrule(lr{.25em}){4-7} \cmidrule(lr{.25em}){8-11} \cmidrule(lr{.25em}){12-15}                                                              

$S_{\text{R}}^{\beta}$                                                                                                                                                  
&49& 49 & 0.00 & 0.00 &13.03& 55.22 &  0.00 &  0.00 & 12.15  & 46.99 &  0.00 &  0.00 & 33.43 &   55.64 \\                                                               
&49& 121& 0.00 & 0.00 &5.39 & 55.23 &  0.00 &  0.00 & 5.35   & 47.13 &  0.00 &  0.00 & 14.47 &  59.66  \\                                                               
&49& 400& 0.00 & 0.00 &1.67 & 58.44 &  0.00 &  0.00 & 1.66   & 45.26 &  0.00 &  0.00 & 3.77  &  70.50  \\                                                               
&121&121& 0.00 & 0.00 &5.38 & 44.49 &  0.00 &  0.00 & 5.97   & 49.79 &  0.00 &  0.00 & 20.83 &  52.79   \\                                                              
&121&400& 0.00 & 0.00 &1.82 & 45.58 &  0.00 &  0.00 & 2.18   & 52.40 &  0.00 &  0.00 & 3.46  &  50.49  \\                                                               
&400&400& 0.00 & 0.00 &2.12 & 39.97 &  0.00 &  0.00 & 2.93   & 37.87 &  0.00 & 0.00  & 1.50  & \multicolumn{1}{c}{$- $} \\                                              
\cmidrule(lr{.25em}){1-3} \cmidrule(lr{.25em}){4-7} \cmidrule(lr{.25em}){8-11} \cmidrule(lr{.25em}){12-15}                                                              

$S_{\text{B}}$                                                                                                                                                          
&49& 49 & 0.00 & 0.00 &13.55 & 50.80 &  0.00 &  0.00 & 12.84 & 44.45 &  0.00 &  0.00 & 31.06  &  46.92    \\                                                            
&49& 121& 0.00 & 0.00 &6.67  & 52.27 &  0.00 &  0.00 & 6.66  & 45.08 &  0.00 &  0.00 & 16.18  &  51.24    \\                                                            
&49& 400& 0.00 & 0.00 &3.22  & 56.22 &  0.00 &  0.00 & 3.20  & 43.69 &  0.00 &  0.00 & 6.79   &  63.46    \\                                                            
&121&121& 0.00 & 0.00 &5.89  & 43.27 &  0.00 &  0.00 & 6.51  & 48.77 &  0.00 &  0.00 & 20.63  &  47.05    \\                                                            
&121&400& 0.00 & 0.00 &2.51  & 44.55 &  0.00 &  0.00 & 3.00  & 51.67 &  0.00 &  0.00 & 4.68   &  49.27    \\                                                            
&400&400& 0.00 & 0.00 &2.36  & 39.41 &  0.00 &  0.00 & 3.62  & 40.78 &  0.00 &  0.00 & 1.67   &  \multicolumn{1}{c}{$- $} \\                                            
\cmidrule(lr{.25em}){1-3} \cmidrule(lr{.25em}){4-7} \cmidrule(lr{.25em}){8-11} \cmidrule(lr{.25em}){12-15}                                                              

$S_{\text{H}}$                                                                                                                                                          
&49& 49 & 0.00 & 0.00 &11.37& 42.69 &  0.00 & 0.00 & 10.92 & 37.88 &  0.00 &  0.00 & 21.33 & 32.77  \\                                                                  
&49& 121& 0.00 & 0.00 &5.83 & 46.05 &  0.00 & 0.00 & 5.86  & 39.78 &  0.00 &  0.00 & 11.91 & 37.01   \\                                                                 
&49& 400& 0.00 & 0.00 &2.87 & 51.70 &  0.00 & 0.00 & 2.88  & 39.37 &  0.00 &  0.00 & 5.33  & 49.21   \\                                                                 
&121&121& 0.00 & 0.00 &5.46 & 40.04 &  0.00 & 0.00 & 5.97  & 45.05 &  0.00 &  0.00 & 15.92 & 35.68   \\                                                                 
&121&400& 0.00 & 0.00 &2.38 & 42.09 &  0.00 & 0.00 & 2.80  & 48.53 &  0.00 &  0.00 & 4.22  & 46.42   \\                                                                 
&400&400& 0.00 & 0.00 &2.29 & 38.33 &  0.00 & 0.00 & 3.05  & 38.90 &  0.00 &  0.00 & 1.64  & \multicolumn{1}{c}{$-$}\\
\bottomrule                                                                                                                                                             
\end{tabular}                                                                                                                                                           
\end{table}                                                                                                                                                            
%\end{landscape}

This analysis reveals that the test based on Kullback-Leibler distance has the best performance with image data.
Small test sizes are important in order to prevent oversegmentation, i.e., the erroneous identification of samples from the same class as coming from different targets.

\subsection{Kullback-Leibler distance robustness}

Limited to the $S_{\text{KL}}$ statistic, the following discussion presents a robustness study of its test size in the presence of outliers.
Only one sample in the test will be contaminated, and the contamination model we adopt consists in allowing each observation from this sample to come from a different distribution than the assumed with a small probability.

The uncontaminated sample is formed by observations from the $\mathcal W(L,\boldsymbol{\Sigma})$ distribution, while the contaminated sample is formed by observations from either this law, with probability $1-\epsilon$, or from $\mathcal W(L,1000\boldsymbol{\Sigma})$ with probability $\epsilon$; in our study, $\epsilon=10^{-5}$.
In other words, the cumulative distribution function of the contaminated sample is given by
$$
\epsilon\cdot F_{\boldsymbol{X}}(\boldsymbol{Z}')+(1-\epsilon)\cdot F_{\boldsymbol{Y}}(\boldsymbol{Z}').
$$  

Table~\ref{table3} shows the empirical test sizes as well as the following additional figures: (c1) the ML estimator mean square error (MSE) for the number of looks, (c2)~the relative mean square error (rMSE) for the covariance matrix estimator adapted from the `total relative bias'~\citep{Cribari-NetoFerrariCordeiro}, (c3) the ratios between mean square errors, and (c4) the distances mean and coefficient of variation.

The mean square error for the number of looks estimation is given by $\operatorname{MSE}(\widehat{L}_I)=\sum_{k=1}^{5500}(\widehat{L}^{(I)}_k-L_I)^2/5500$, where $\widehat{L}^{(I)}_k$ represents the obtained estimates at the \textit{k}th Monte Carlo replication for population $I\in\{X,Y\}$.         
For the covariance matrix, the relative mean square error is 
$$
\operatorname{rMSE}(\widehat{\boldsymbol{\Sigma}}_I)=\frac1{5500}\sum_{k=1}^{5500} \sum_{h=1}^3\frac{((\widehat{\boldsymbol{\Sigma}}^{(I)}_k)_{hh}-(\boldsymbol{\Sigma}_I)_{hh})^2}{( \boldsymbol{\Sigma}_I)_{kk}},
$$
where $\widehat{\boldsymbol{\Sigma}}^{(I)}_k$ is the estimate of the covariance matrix at the \textit{k}th Monte Carlo experiment for the population $I$.
The ratios for these measures are denoted by 
$$
r_1=\frac{\operatorname{MSE}(\widehat{L}_X)}{\operatorname{MSE}(\widehat{L}_Y) }
\text{ and } 
r_2=\frac{\operatorname{rMSE}(\widehat{\boldsymbol{\Sigma}}_X)}{\operatorname{rMSE}(\widehat{\boldsymbol{\Sigma}}_Y)}.
$$

\begin{table}[hbt]                                                                                               
\centering                                                                                                        
%\scriptsize                                                                                                       
\caption{Robustness for forest-1.
Mean square errors scaled by 100} \label{table3}
%\begin{tabular}{ccc rrrrccrccr}
\begin{tabular}{c@{ }c@{ }c  r@{\quad}r@{\quad}r@{\quad}r@{\quad}r@{\quad}r@{\quad}r@{\quad}r@{\quad}r@{\quad}r}
\toprule
$N_X$ & $N_Y$ & $L$ & \rotatebox{90}{$1\%$} & \rotatebox{90}{$5\%$} & \multicolumn{1}{c}{\rotatebox{90}{$\overline{d}$}}&  \multicolumn{1}{c}{\rotatebox{90}{CV}}
& \multicolumn{1}{c}{\rotatebox{90}{$\operatorname{MSE}$($\widehat{L}_X$)}} & \multicolumn{1}{c}{\rotatebox{90}{$\operatorname{MSE}$($\widehat{L}_{Y}$)}} & 
\multicolumn{1}{c}{\rotatebox{90}{$r_1$}} & \multicolumn{1}{c}{\rotatebox{90}{$\operatorname{rMSE}$($\widehat{\boldsymbol{\Sigma}}_X$)}} & 
\multicolumn{1}{c}{\rotatebox{90}{$\operatorname{rMSE}$($\widehat{\boldsymbol{\Sigma}}_Y$)}}&
\multicolumn{1}{c}{\rotatebox{90}{$r_2$}} \\ 
\cmidrule(lr{.25em}){1-3} \cmidrule(lr{.25em}){4-13} 
49 & 49  & 4   & 1.82	&	6.87	&	10.60	 &	45.48	&  16.28	&	18.72	 &	0.87 & 55.36	&   55.52		& 1.00 \\
   &     & 8   & 0.95	&	4.62	&	 9.73	 &	46.66	&  14.75	&	16.96	 &	0.87 & 27.78	&   28.21		& 0.98 \\
   &     & 16  & 0.84	&	4.07	&	 9.41	 &	47.00	&  14.13	&	16.03	 &	0.88 & 13.70	&   14.67		& 0.93 \\
49 & 121 & 4   & 2.09	&	7.80	&	10.71	 &	46.06	&  16.23	&	17.54	 &	0.93 & 55.05	&   23.74		& 2.32 \\
   &     & 8   & 1.02	&	4.49	&	 9.64	 &	47.20	&  14.80	&	15.41	 &	0.96 & 27.50	&   11.96		& 2.30 \\
   &     & 16  & 0.85	&	4.07	&	 9.44	 &	46.67	&  14.40	&	14.71	 &	0.98 & 13.50	&   6.56		& 2.06 \\
49 & 400 & 4   & 1.67	&	8.07	&	10.79	 &	45.20	&  16.31	&	16.92	 &	0.96 & 54.53	&   7.85		& 6.94 \\
   &     & 8   & 1.07	&	5.00	&	 9.77	 &	46.95	&  14.83	&	14.76	 &	1.00 & 27.05	&   4.41		& 6.13 \\
   &     & 16  & 0.56	&	4.09	&	 9.55	 &	45.57	&  14.21	&	13.97	 &	1.02 & 13.74	&   2.70		& 5.08 \\
121& 121 & 4   & 2.04	&	7.64	&	10.75	 &	45.86	&  14.90	&	17.59	 &	0.85 & 22.24	&   23.26		& 0.96 \\
   &     & 8   & 0.80	&	4.58	&	 9.72	 &	45.40	&  13.45	&	15.44	 &	0.87 & 10.95	&   12.22		& 0.90 \\
   &     & 16  & 0.96	&	4.44	&	 9.56	 &	46.69	&  12.87	&	14.71	 &	0.87 & 5.60		&   6.60		& 0.85 \\
121& 400 & 4   & 2.15	&	7.69	&	10.84	 &	45.84	&  14.88	&	16.97	 &	0.88 & 22.08	&   7.78		& 2.84 \\
   &     & 8   & 1.05	&	4.87	&	 9.88	 &	46.04	&  13.30	&	14.75	 &	0.90 & 11.06	&   4.41		& 2.51 \\
   &     & 16  & 1.62	&	5.44	&	 9.73	 &	48.89	&  12.86	&	13.97	 &	0.92 & 5.52		&   2.73		& 2.02 \\
400& 400 & 4   & 1.93	&	7.87	&	10.91	 &	44.79	&  14.31	&	16.91	 &	0.85 & 6.54		&   7.88		& 0.83 \\
   &     & 8   & 1.22	&	5.96	&	10.12	 &	46.00	&  12.65	&	14.71	 &	0.86 & 3.33		&   4.40		& 0.76 \\
   &     & 16  & 1.22	&	6.27	&	10.22	 &	46.57	&  12.17	&	14.01	 &	0.87 & 1.70		&   2.74		& 0.62 \\
\bottomrule
\end{tabular}
\end{table}

The results presented in Table~\ref{table3} reveal that the mean square errors are reduced when larger windows or number of looks are considered.
This behavior is expected, since in those cases the signal-to-noise ratio is improved.

For  situations where $N_X=N_Y=N\in \{49,121,400\}$, the mean square errors for the estimators of the number of looks and covariance matrix increase when contamination is introduced. 
Additionally, increasing the value of $N$ does not significantly affect the ratio $r_1$.
Simultaneously, ratio $r_2$ decreases.
This last fact is justified by the contamination model nature, and indicates that the contamination effect on the empirical test size is strongly related to the estimation of the covariance matrix.

\begin{figure}[htb]
\centering
\includegraphics[width=1\linewidth]{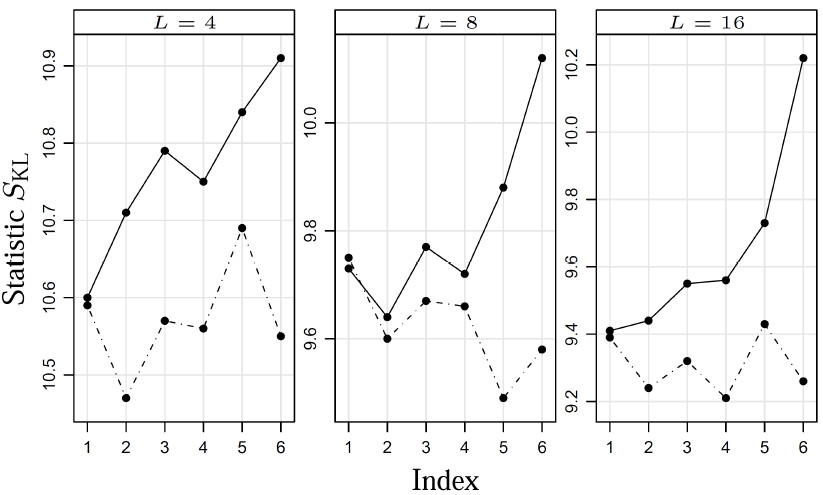}
\caption{Stochastic distances with $\epsilon=\{0,10^{-5}\}$. For each sample size, scenarios were lexicographically ordered: $[i:(i;N_X,N_Y)=\{(1;49,49),$ $(2;49,121),$ $(3;49,400),$ $(4;121,121),$ $(5;121,400),$ $(6;400,400)\}]$. } 
\label{distcontamin}
\end{figure}

Figure~\ref{distcontamin} plots $S_{\text{KL}}$ in pure and contaminated scenarios.
As expected, the contamination effect under $S_{\text{KL}}$ increases when considering situations with larger windows.
However, this effect is diminished when $L$ increases.

\section{Conclusion}~\label{ChJS:sec4}

This work aimed to provide a statistical formalism for the use of information theory tools in the understanding of PolSAR imagery.
Five novel hypothesis tests for polarimetric speckled data were assessed.
Their associate statistics were based on chi-square, Kullback-Leibler, R\'enyi, Bhattacharyya, and Hellinger distances between ML estimators for the parameters that index the scaled multilook complex Wishart law.
As a comparison criterion, empirical test sizes were quantified in several situations, which included pure and contaminated data.
An application to actual data was performed.  

The test based on the Kullback-Leibler distance had the closest empirical size to the nominal level.
Under contamination, the ML estimator for the covariance matrix was significantly affected, yielding the $S_\text{KL}$ statistic sensitive to outliers. 
However, this effect was reduced increasing the number of looks, i.e., this contrast measure presented better performance in the presence of contamination when signal-to-noise ratio was controlled.
With the exception of the chi-square measure, all stochastic distances presented good performance when applied to a real PolSAR image.

Many venues for research are still open in PolSAR image modelling and analysis.
Using stochastic distances in actual segmentation and classification algorithm requires fast and reliable inference procedures which, ideally, are resistant (robust) to plausible contamination.
Expressive and tractable models for spatial correlation under the Wishart law, and their implication in inference is another promising research direction.
Extending the results here presented to more general models than the scaled multilook complex Wishart law, for instance to the polarimetric $\mathcal G^0$ and $\mathcal G^H$ distributions, or even to the still more general complete polarimetric $\mathcal G$ model \citep[see][]{FreitasFreryCorreia:Environmetrics:03,PolarimetricSegmentationBSplinesMSSP} would greatly enhance the capability of understanding this class of images.

\section*{Acknowledgements}

The authors are grateful to CNPq, Fapeal and Facepe for their support to this work.
Afterword

\section*{Afterword}

This invited paper is an extended and detailed version of a talk given at the III Simposio de Estad\'istica Espacial y Modelado de Im\'agenes~(SEEMI), 
that took place at Foz do Igua\c{c}cu, Brazil, in December 2010. 
Though many results discussed here have been already published in specialized journals and conferences, in this presentation we intend to achieve
the wide audience of probability and statistics, with a bias towards applications in signal and image processing.

%\bibliographystyle{agsm}
%\bibliography{../bibtexart}

\end{document}